# Cypher4BIM: Releasing the Power of Graph for Building Knowledge Discovery

**Junxiang Zhu** [1, 4, *], **Nicholas Nisbet** [2], **Mengtian Yin** [1], **Ran Wei** [3], **Ioannis Brilakis** [1]

1. Department of Engineering, University of Cambridge, Cambridge, United Kingdom
2. Bartlett School of Sustainable Construction, University College London, London, United Kingdom
3. School of Computing and Communications, Lancaster University, Lancaster, United Kingdom
4. School of Design and the Built Environment, Curtin University, Bentley, Australia
* Correspondence: jz652@cam.ac.uk

**Abstract:** Graph is considered a promising way for managing building information. A new graphic form of IFC (Industry Foundation Classes) data has just been developed, referred to as IFC-Graph. However, understanding of IFC-Graph is insufficient, especially for information query. This study aims to explore graphic building information query and develop a graph query language tailored for IFC-Graph. A series of tasks were carried out, including a) investigating the structure of IFC data and the main types of information in IFC, b) investigating the graph query language Cypher, and c) developing a set of tailored functional query patterns. The developed language is referred to as Cypher4BIM. Five IFC models were used for validation, and the result shows that Cypher4BIM can query individual instances and complex relations from IFC, such as spatial structure, space boundary, and space accessibility. This study contributes to applications that require effective building information query, such as digital twin.

## 1 Introduction

This paper explores an innovative way for building information query and knowledge discovery by using graphs. Building information query refers to retrieving explicit information in digital building information models, and building knowledge discovery refers to identifying and retrieving implicit information in building information models. Implicit information is information that cannot be directly retrieved, such as the spatial structure and element connectivity. The problem that this paper addresses is understanding how to effectively query information from a new graphic form of building information. This problem is important, because building information query is the very first step of utilising building information (e.g., from visualisation to analysis), and effective building information query contributes to building knowledge discovery. If not properly handled, it can affect the performance of the whole system [1,2].

Digitalisation is a trend over the world across disciplines. The geospatial industry is digitalising the natural environment, the Architecture, Engineering, Construction, and Operation (AECO) domain is digitalising the built environment via building information modelling (BIM) [3-5], and the manufacturing industry has been digitalising their products via computer aided design (CAD). These areas have united in the field of digital twin, especially the geospatial industry and the AECO domain. They are trying to create digital representations of the natural environment and built environment to support evidence-informed decision-making [6-8]. While digitalisation brings benefits to effective decision making, one of the challenges coming along is the efficient query of digital data. This is an outstanding problem for areas with multiple stakeholders such as the construction industry that use heterogeneous data [9]. With the digitalisation of construction projects, practitioners are faced with the difficulty of efficiently finding the information they need. For example, it is estimated that in Australia, where the construction industry is the third largest industry, construction workers spend about 4.9 hours hunting down project data each week, resulting in a loss of $15.6 billion in labour costs [10].

The challenge in efficient information query in the construction industry comes from, as stated above, the use of heterogeneous data that are in various formats and forms (e.g., individual files and isolated databases). This leads to the well-known interoperability issue in the AECO domain, i.e., data created by one system cannot be effectively used by another [11,12]. In recent years, with the advance of data science where new data formats and new database systems have been developed, it is now possible to manage the interoperability issue, in terms of both syntactic interoperability and semantic interoperability. Syntactic interoperability means data created by one system can be correctly read by other systems, while semantic interoperability means the data can be correctly understood by other systems [13]. Building information created by one system (e.g., a BIM authoring tool) can now be read and correctly interpreted by other systems within AECO. However, it is still a challenge to effectively query information from these digital building models due to their interconnected structure that is full of relationships [14] and the growing size of Industry Foundation Classes (IFC) [15]. This challenge is even obvious in the broader context of digital twins [16,17] and smart cities [8,18,19]. In short, the complexity of the data structure makes it difficult to systematically extract information from IFC building models [20].

One of the promising techniques to address this issue is graph technology, which can effectively handle relationships [21,22]. It is expected that by converting digital building models into graphs, information in the digital models can be effectively retrieved by using advanced graph traversal algorithms, even the implicit hidden information [1]. There were a few initial attempts by researchers trying to reveal the power of graph in building information query, but they were all stranded in the creation of a full graphic representation of building information that is compliant with the original IFC data model. Therefore, their assessments of the use of graph in building information query are largely partial, which impairs the further step of graph-based building knowledge discovery. Zhu et al. [1] developed a method to fully convert building information into labelled property graph (LPG) that is compliant with the original IFC data model, referred to as IFC-Graph. This advance enables a full assessment of the graph technology in facilitating building information query, and even building knowledge discovery. Therefore, this study tries to reveal the power of graph on building information query by developing a graph-based language for querying graphic building information.

The remainder of this paper is organised as follows. Section 2 introduces the existing methods for building information query. Section 3 presents the development of the graph-based language for building information query. Section 4 is about the validation of the query language by using digital building models. Discussions are presented in Section 5, and Section 6 conclusions the paper.

**2 Literature review**

*2.1 Building information query*

Information query refers to the process of finding the required information from datasets, and a query is a statement requesting the retrieval of information [23]. After reviewing literature, the existing methods for building information query are summarised in Figure 1. In general, there are two approaches for querying building information, depending on the environments used, including file-processing environment and database environment.

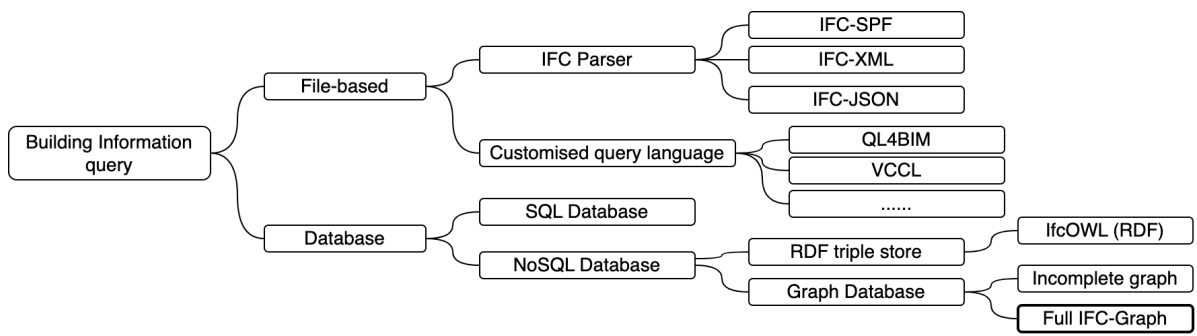

Figure 1 Existing methods for building information query.

2.1.1 The file-based environment

File-based environment queries building information from individual files that contains digital models, such as IFC-SPF (STEP Physical File). There are two subtypes for this environment. The first is using existing file parsers, such as IfcOpenShell [24] and IfcEngine DLL [25] for IFC-SPF and the general XML (Extensible Markup Language) parser, XQuery [26], for IFC-XML [9]. Parsers provide capability in effective building information access, but not efficient building information query. As stated by Silberschatz et al. [23], conventional file-processing environments do not allow a convenient and efficient way for needed data to be retrieved. However, this is the most commonly used way for building information query, mainly because IFC is currently optimised for file-based exchange [15].

The second is customised query languages, such as the Query Language for Building Information Models (QL4BIM) by Daum and Borrmann [27] and the Visual Code Checking Language (VCCL) by Preidel et al. [28]. Some earlier IFC-based query languages include the Express Query Language (EQL), the Partial Model Query Language (PMQL), the Generalised Model Subset Definition (GMSD), and BimQL [20]. These query languages are domain specific. For example, QL4BIM is for querying spatial topology of building elements by processing and analysing geometric information (i.e., boundary representation), while VCCL provides a visual way to conduct code compliance check for produced building models.

Apart from these two subtypes, an even earlier way for 'information query' was through schema mapping. By defining a mapping schema between two data models using languages such as EXPRESS-X [29], the required information in the original data model is filtered and stored in the new data model. The data in the new data model is then the 'query' result.

2.1.2 The database environment

Compared with file-based environments, database environments can provide better support for information query. Database management system (DBMS) are specifically designed for data storage, data query, and transaction management [23]. They can perform efficient information query by using well-designed query languages, such as the SQL (Structured Query Language) for relational databases and the graph query languages (such as Cypher) for graph databases [30,31]. This advantage of databases has inspired buildingSMART, i.e., the organisation managing the IFC standard, to develop a query language and an object-based API (application programming interface) for IFC when planning for IFC5 [15].

Among the two general types of DBMS (i.e., relational and NoSQL, or Not Only SQL), the previous studies mainly used NoSQL database management system, especially graph database, to handle building information. This is because the conventional SQL DBMS cannot effectively handle relationships [21], thus not efficient query as well. In contrast, graph database is tailored for handling relationships, such as the friend-of-a-friend relationship in social network [32]. Such a capability of graph can also be used to reveal the hidden relation between building elements.

*2.2 Graph database and Resource Description Framework*

RDF (Resource Description Framework), which is for constructing the Semantic Web, is also considered to be graph based [33-35]. RDF can be used in file-based environments and database management environments. A text-based RDF file (e.g., in Turtle format) can be queried using regular expression, and after being put into a database, they can be queried using SPARQL (i.e., the native query language for RDF) [36,37] or graph query language. Several studies have compared RDF with LPG (or graph database) in terms of data query efficiency, such as Donkers et al. [38] and Alocci et al. [39]. However, this paper tries to argue that RDF should not be compared with graph database in terms of data storage and query, because, in short, they are different technologies, even though they are both related to data.

First, graph database is database technology, while RDF is above database technology for data linking. As stated by Antoniou et al. [40], semantic web uses the existing infrastructure of the web, which includes the database technology. Second, database technology is focusing on data storage, data query, and transaction management, while RDF is focusing on inter-domain data linking, which relies on database technology for implementation. Third, database technology is using the closed world assumption, while RDF is using the open world assumption [41]. A detailed discussion about the differences and connections between graph database and RDF has been provided in [1,11]. What researchers are excited about semantic web, or RDF, is the final use of reasoners to find relations (hidden information) from RDF data through machine inferencing. Such a capability in inferencing can be realised by graph technology as well [21].

This study limits graph database to **native** graph database which is defined by Robinson et al. [21] that uses native graph storage (i.e., using labelled property graph, or LPG, as data storage structure) and uses native graph processing (i.e., using index-free adjacency). The RDF Triplestores, i.e., databases used to store RDF triples [42], are not within this scope.

*2.3 Challenges in graph-based building information query*

2.3.1 The lack of a full graphic representation of building information

While it is clear that graph database using LPG provides a good capability in handling relationships and querying graphs, the problem was that there was no full graphic representation of building data available for conducting a full assessment, due to the lack of a method to fully and automatically convert IFC data into a graph structure that is compliant with the original IFC data model, or IFC-compliant LPG [1]. There were studies trying to convert IFC data into LPG. For example, Ali et al. [43,44] extracted information from IFC-SPF and converted that into LPG. They examined the use of graph database in storing building information and the use of graph-based query for pathfinding. However, methods developed by previous studies for data conversion and graph generation was largely partial, just right to meet their own information needs for specific use cases, such as data analysis and visualisation [45-48], they did not convert IFC data into IFC-compliant LPG. The workflow for converting and using LPG-based building information by previous studies is presented in Figure 2. IFC-SPF is represented by nodes and edges with dashed lines, because IFC data can conceptually thought of as a graph, or more precisely, an implicit graph, while explicit graph is represented by nodes and edges with solid lines.

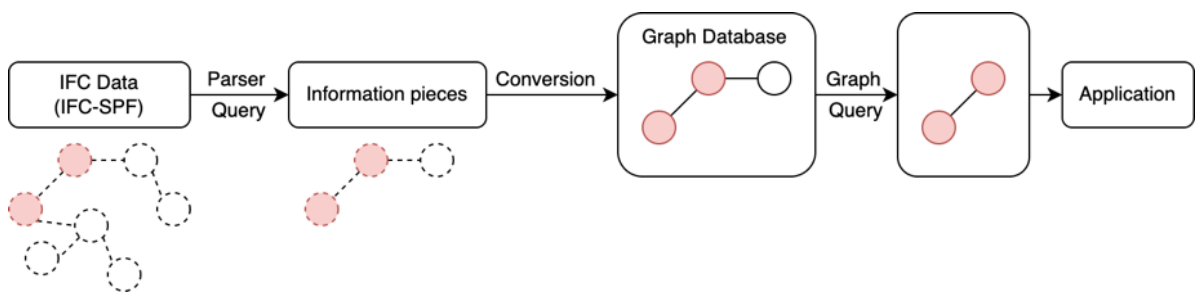

Figure 2 Graph generation and graph query by previous studies.

2.3.2 The lack of a general-purpose, graph-based query language

The above problem has been addressed by Zhu et al. [1]. They developed an algorithm to fully convert IFC-SPF files into IFC-compliant LPG. Such LPGs are referred to as IFC-Graph, they enable the direct retrieval of information from graph. IFC-Graph is a new form of IFC data, and many aspects have not been sufficiently investigated, such as a) what is the difference between querying general graphic data and querying IFC-Graph? b) what information can be queried from IFC-Graph and what information can be queried more efficiently from IFC-Graph? b) what is the best way to query building information from IFC-Graph? These questions are critical for effectively retrieving information from IFC-Graph, or more broadly, from any graphic building information. In addition, while there are general graph query languages available, such as Cypher [31] and GraphX [49], a language more specific to graphic building information is needed, as information query is closely related to the underlying data model [20].

## 3 Cypher4BIM - a graph-based building information query language

Based on the above identified problems, a query language was developed in this study, referred to as Cypher4BIM. This query language is based on the general-purpose graph query language (GQL), Cypher, which is developed by Neo4j. The relationship between Cyphter4BIM and Cypher is like the relationship between Python packages and Python. Python is the hosting language, while packages are based on the hosting language but provide more advanced functionalities.

The joint use of graphic building information and graph query language can create a new paradigm of using building information, as presented in Figure 3. To develop Cypher4BIM, the aspects of IFC and Cypher presented in the following sections have been thoroughly investigated.

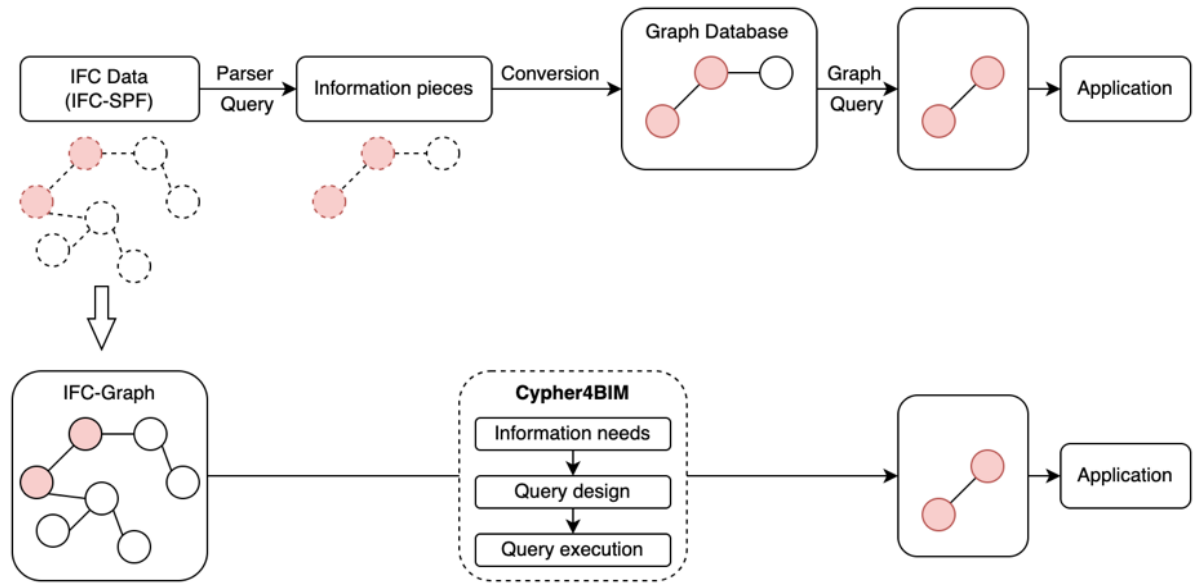

Figure 3 Graph-based IFC information query via Cypher4BIM.

### 3.1 IFC data model (structure) and IFC data

A good understanding of the underlying data model, or the structure of IFC, is essential for effective information query [20]. The relationship between IFC data model and IFC data is that the IFC data model is a higher level of abstraction of IFC data, which provides useful reference when querying IFC data.

In the IFC data model, the most basic elements are entities, which are used to define objects (IfcObjectDefinition), properties (IfcPropertyDefinition), and relationships (IfcRelationship). Object entities define types of elements, their attributes, while relationship entities define the relations between object entities and/or property entities. A detailed description of the IFC data model can be found in [50]. The relationship entities in IFC are like bridges connecting entities, implicitly making the IFC data model a graph. Due to the existence of the various relationship entities in IFC, many useful types of information are available in IFC data, such as property sets, connectivity between building elements, accessibility to space, space boundary, and the spatial structure.

### 3.2 IFC-Graph (LPG-based IFC data)

IFC-Graph is a full graphic representation of IFC data. In the study by Zhu et al. [1], the IFC elements (i.e., entities and attributes) are mapped to graph elements (i.e., nodes, edges, properties) using the mapping presented in Table 1. To distinguish the attributes in IFC, the concepts of intrinsic attribute and extrinsic attribute were borrowed from model-driven engineering (MDE) [51]. Intrinsic attributes are attributes having only primitive data values, while extrinsic attributes point to other instances. Entities are mapped to nodes, intrinsic attributes are mapped to properties of nodes, and extrinsic attributes are mapped to edges. An advantage of using such a mapping is the preservation of the original IFC data structure, which can facilitate graph-based building information query.

Table 1 Mapping IFC elements (entities and attributes) to graph elements (nodes, edges, properties).

|  | From (IFC) | To (IFC-Graph) |
|---|---|---|
| **Entity** | Objects | Nodes |
|  | Relationships | Nodes |
|  | Property sets | Nodes |
| **Attributes** | Intrinsic attributes | Properties of nodes |
|  | Extrinsic attributes | Labels of edges |

*3.3 Graph-based building information query*

The first step of information query is to determine information need, which is closely related to the use case. This study tries to provide a generic way for information query and has no assumption on the use case. Therefore, instead of focusing on what information is needed from IFC, this study looks at the problem from the other side, i.e., what information can be provided by IFC.

3.3.1 The graph database and query language

As stated above, database environments provide better support for data query. In this study, IFC-Graphs were stored in neo4j, a native graph database management system that uses LPG for data storage and native graph processing for data query [21]. Cypher is the native query language for neo4j. One important reason for this study to use Cypher as a hosting language is its close relation with the newly published international standard for graph query language, i.e., ISO/IEC 39075:2024 [52,53]. In addition, the open-source implementation of Cypher, i.e., openCypher, has been adopted by other database management systems, such as ArcadeDB, Katana Graph and Memgraph [54].

A key concept in graph query is pattern. Patterns are used to represent human's knowledge of an area or subject when retrieving information. They have been used in regular expression for the extraction of information from text [55]. In regular expression, patterns can be used to extract certain texts, for example, starting or ending with specific letters. Patterns in graph query are more powerful than that and can be used to retrieve hidden information. That is the reason graph query language is referred to as inferencing query language by Robinson et al. [21].

3.3.2 The basic query patterns for general graphs

Table 2 presents the basic patterns for querying nodes, relationships, and paths using Cypher. In a query pattern, nodes are indicated by '( )', which can include labels and properties. Labels are following a ':' and properties are within a pair of curly brackets. Edges are denoted by '− −', which can include labels as well, and arrows can be used to indicate the direction of the connection. Paths are formed by a series of nodes and edges. A more detailed explanation for the basic patterns can be found in [56].

Table 2 The basic patterns for querying nodes, relationships, and paths from graph.

|  | Type | The basic pattern | Query |
|---|---|---|---|
| 1 | Nodes | $(n: Label\{property: value\})$ | $match\ (n: Label\{property: value\})\ return\ n$ |
| 2 | Edges | $<-[r: Label]->$ | $match\ (n1) <-[r: Label]-> (n2)\ return\ r$ |
| 3 | Paths | $(n1) <-[r]-> (n2)$ | $match\ p = (n1) <-[r]-> (n2)$ return p |

3.3.3 The functional query patterns for IFC-Graph

The basic query patterns are not the best way to query IFC-Graph. This is because the real 'relation' in IFC-Graph is indicated by the objectified relationships that have been mapped to nodes, while general graphs use edges to indicate relation (see Figure 4). Adjustment to the basic query patterns is thus required to suit the IFC data model. Based on the basic query patterns, a set of advanced query patterns were further developed for querying entities and relations from IFC-Graph, referred to as functional query patterns, as shown in Table 3.

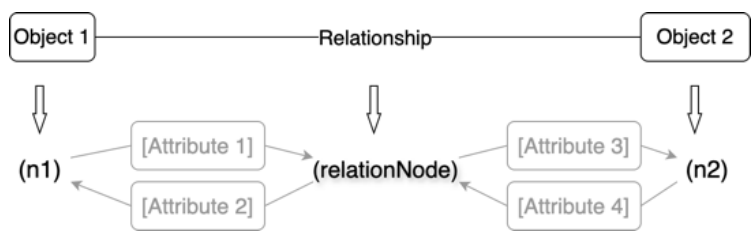

Figure 4 Relation between the basic patterns and the functional patterns.

Table 3 The functional patterns for querying entities and relations from IFC-Graph.

| | Information in IFC-Graph | The functional pattern |
|---|---|---|
| 1 | Entities | $(n: IfcClass\{attribute: value\})$ |
| 2 | Relations | $p = (node1) --(relationNode) --(node2)$ |

The functional query patterns are variant of the basic query patterns, and there are some differences. First, the type of entity (IfcClass) is used to replace the general 'Label', enabling it to retrieve instances of specific entities (objects, properties, and relationships). Second, for retrieving relations, labels of edges are not used, because the real relationship is stored in nodes. In IFC-Graph, IFC relationship entities are mapped to nodes, this is because objectified relationship entities have their own intrinsic attributes, such as the PhysicalOrVirtualBoundary attribute of IfcRelSpaceBoundary [57]. If mapped to edges, these intrinsic attributes will be lost. Therefore, the functional pattern for relation query is $(node1) - [attribute1] - (relationNode) - [attribute2] - (node2)$, where [attribute] can be omitted, making it '$(node1) --(relationNode) --(node2)$'. It is interesting to notice that the functional query pattern for relations has a similar structure to the subject-predicate-object structure used by RDF [40].

### 3.3.4 Building knowledge expressed by graph patterns

With the functional patterns, human knowledge about buildings/assets can be expressed and used to query IFC data. For example, it is a common sense that a room (or space in IFC) can be surrounded by floors, walls, windows, doors, and columns. Such knowledge can be expressed by the pattern presented in Listing 1.

Listing 1 Pattern for representing knowledge of space boundary.

---

$(: IfcSpace) --(: IfcRelSpaceBoundary) --(: IfcSlab)$ OR $(: IfcSpace) --(: IfcRelSpaceBoundary) --(: IfcWall)$ OR $(: IfcSpace) --(: IfcRelSpaceBoundary) --(: IfcWindow)$ OR $(: IfcSpace) --(: IfcRelSpaceBoundary) --(: IfcDoor)$ OR $(: IfcSpace) --(: IfcRelSpaceBoundary) --(: IfcColumn)$

---

### 3.3.5 Cypher4BIM queries

Based on the knowledge about buildings and the IFC standard, a set of functional Cypher4BIM queries were developed. The knowledge required and the involved IFC entities are presented in Table 4, and the corresponding queries are presented in Table 5. By doing this, human knowledge about buildings is represented by patterns and can be directly used in building information query from IFC-Graph. In this paper, the query results are returned as paths. Once a path is retrieved, so as the nodes and edges contained in the path.

Table 4 Knowledge and key IFC entities involved in functional queries.

| Type | Information to be queried | Knowledge about building and IFC | Key IFC entities involved |
|---|---|---|---|
| Entity | 1) Instance with a specific ID | Instances have a unique ID to uniquely identify them. | Any |
| | 2) Instance(s) with specific attribute value | Instances have attributes, which can be used to filter instances. | Any |
| | 3) Instances of a specific entity | An entity (class) can have multiple instances; A superclass can have multiple subclasses; it is possible to use the superclass to retrieve the instances of its subclasses. | Any |
| | 4) Instances meeting specific conditions | Conditions (e.g., <, >, =) can be used to filter instances | Any |
| Relation | 5) Spatial structure | IFC has a spatial structure comprising spatial structure elements, which is IfcSite → IfcBuilding → IfcBuildingStore → IfcSpace, in this structure, an upper level can contain multiple instances of the lower level. IfcSite is connected to IfcProject. | IfcRelAggregates IfcSpatialStructureElement |
| | 6) Spatial containment | The spatial structure elements (IfcSite, IfcBuilding, IfcBuildingStorey, and IfcSpace) can accommodate building elements, e.g., walls, doors, columns, etc. | IfcRelContainedInSpatialStructure IfcSpatialStructureElement IfcBuildingElement |
| | 7) Space boundary | Spaces (or rooms) are bounded by building elements, such as doors, windows, slabs, columns, and walls. | IfcRelSpaceBounday IfcBuildingElement |
| | 8) Space accessibility | Rooms are connected by doors, and in IFC, the virtual element is used to connect virtually separated space. | IfcSpatialStructureElement IfcRelSpaceBounday IfcBuildingElement IfcSpace |
| | 9) Connectivity | Building elements, such as walls, are connected to each other. | IfcRelConnectsPathElements IfcBuildingElement |
| Property and quantity | 10) Property sets | Building elements can have properties stored in property sets. | IfcRelDefinesByProperties IfcBuildingElement IfcPropertySet |
| | 11) Quantity sets | Building elements can have quantities stored in quantity sets. | IfcRelDefinesByProperties IfcBuildingElement IfcQuantitySet |

Table 5 Functional Cypher4BIM queries.

| | Building information | Pattern/Query |
|---|---|---|
| 1) | Instance with a specific id | match (n{id: id}) return n<br>// Get the instance with an id of 121<br>match (n{id: 121}) return n |
| 2) | Instance(s) with specific attribute value | match (n: {attribute: value}) return n |
| 3) | Instances of a specific entity | match (n: IfcClass) return n<br>// Get all instances of IfcDoor<br>match (n: IfcDoor) return n |
| 4) | Instances meeting specific conditions | // Get doors with a width larger than 0.9 m<br>match (n: IfcDoor) where n.OverallWidth > 0.9<br>return n |
| 5) | Spatial structure | // Get IfcSite, IfcBuilding, IfcBuildingStorey, IfcSpace and their relations<br>match p = (n1: IfcSpatialStructureElement) −−(r: IfcRelAggregates) −−(n2: IfcSpatialStructureElement)<br>return p |
| 6) | Spatial containment | // Get building elements in spatial structure elements<br>match p = (n1: IfcBuildingElement) −−(r: IfcRelContainedInSpatialStructure) −−(n2: IfcSpatialStructureElement) return p |
| 7) | Space boundary | // Get spaces and building elements bounding them<br>match p = (n1: IfcSpace) −−(r: IfcRelSpaceBoundary) −−(n2: IfcBuildingElement)<br>return p |
| 8) | Space accessibility | // Get spaces connected by doors and virtual elements<br>match p = (n1: IfcSpatialStructureElement) −−(r: IfcRelSpaceBoundary) −−(n2)<br>where n2.name in ['IfcDoor', 'IfcVirtualElement']<br>return p |
| 9) | Connectivity | // Get walls connected to each other<br>match p = (n1: IfcWall) −−(r: IfcRelConnectsPathElements) −−(n2: IfcWall) return p |
| 10) | Property Set | // Get all building elements and their property sets<br>match p = (n1: IfcBuildingElement) −−(r: IfcRelDefinesByProperties) −−(n2: IfcPropertySet) return p |
| 11) | Quantity Set | // Get all building elements and their quantity sets<br>match p = (n1: IfcBuildingElement) −−(r: IfcRelDefinesByProperties) −−(n2: IfcQuantitySet) return p |

## 4 Validation

*4.1 Data and methods*

Five IFC building models, as listed in Table 6, were used to validate the developed queries, assessing their performance. These IFC models were converted into IFC-Graphs using the method developed by Zhu et al. [1]. The IFC models and their corresponding IFC-Graphs are presented in Figure 5. Scripts have been developed to carried out the queries of individual instances and relations (See Appendix).

Table 6 IFC models used in this study for validation.

|   | Model | IFC-SPF | | | IFC-Graph | |
|---|---|---|---|---|---|---|
|   |   | File Size | Entities/Instances | Version | Nodes | Edges |
| 1 | Duplex | 1.7 MB | 162/27529 | IFC2x3 | 28729 | 58334 |
| 2 | House | 2.6 MB | 189/44249 | IFC4 | 48786 | 79137 |
| 3 | Smiley | 6.1 MB | 168/110176 | IFC4 | 113903 | 180907 |
| 4 | Institute | 10.9 MB | 189/147712 | IFC4 | 194845 | 285504 |
| 5 | Clinic | 13.2 MB | 170/209148 | IFC2x3 | 216633 | 496434 |

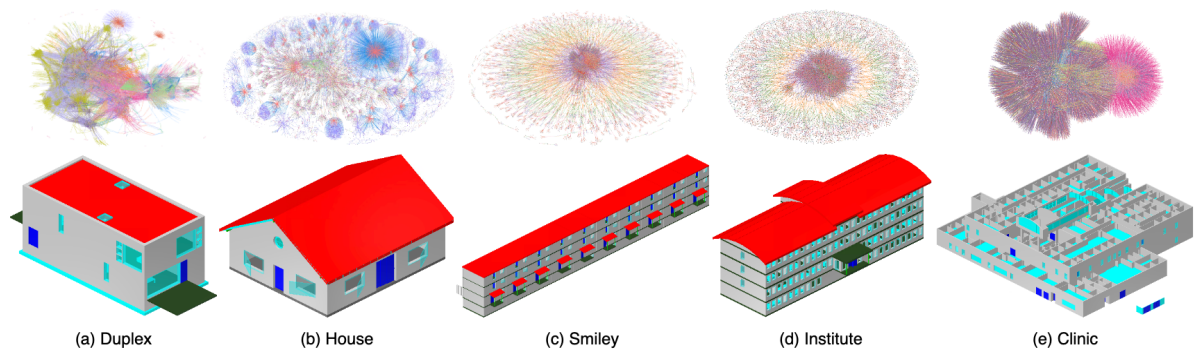

Figure 5 IFC models and IFC-Graphs used for validation.

*4.2 Retrieving instances of entities*

Instances of entities were queried using both IfcOpenShell (within Python) and Cypher4BIM. The results were compared to check if Cypher4BIM can retrieve the same instances as IfcOpenShell. The codes for querying are listed in Table 7. The queried entities include IfcObjectDefinition, IfcRelationship, and IfcPropertyDefinition, i.e., the immediate child entities of IfcRoot.

Table 7 Queries used for validation.

|   | Tool | Query |
|---|---|---|
| 1 | IfcOpenShell | # Find all the instances of IfcClass in IFC-SPF and return the number of instances<br>function get_instance(IfcClass): n = ifcopenshell.by_type(IfcClass) return len(n) |
| 2 | Cypher4BIM | # Find all the instances of IfcClass in IFC-Graph and return the number of instances<br>match (n: IfcClass) return count(n) |

The query results are presented in Table 8. By comparing the number of retrieved instances, it is certain that Cypher4BIM can retrieve the same individual instances of entities as IfcOpenShell. In addition, it is obvious that the query time of IfcOpenShell is dependent on the total number of instances in a building model and the number of instances to be returned. For instance, the queries for Object, Relationship, and Property from the house model returned 146, 659, and 499 instances within 0.291 ms, 1.311 ms, and 1.032 ms, respectively. In the meanwhile, the query time of Cypher4BIM is relatively stable, regardless of the number of instances to be returned. It is thus reasonable to claim that when dealing with large building models, Cypher4BIM can provide higher query efficiency.

Table 8 Query results of individual instances of entities.

|  | Model | Entity | Number of Queried Instances | | Query Time (milliseconds, ms) | |
|---|---|---|---|---|---|---|
|  |  |  | IfcOpenShell | Cypher4BIM | IfcOpenShell | Cypher4BIM |
| 1 | Duplex | Object | 333 | 333 | 2.400 | 2.212 |
|  |  | Relationship | 2068 | 2068 | 18.934 | 2.262 |
|  |  | Property | 1492 | 1492 | 15.363 | 2.155 |
| 2 | House | Object | 146 | 146 | 0.291 | 3.018 |
|  |  | Relationship | 659 | 659 | 1.311 | 2.871 |
|  |  | Property | 499 | 499 | 1.032 | 2.923 |
| 3 | Smiley | Object | 1297 | 1297 | 0.729 | 2.357 |
|  |  | Relationship | 5631 | 5631 | 7.211 | 2.074 |
|  |  | Property | 2625 | 2625 | 5.850 | 2.044 |
| 4 | Institute | Object | 1210 | 1210 | 5.281 | 1.971 |
|  |  | Relationship | 6374 | 6374 | 22.185 | 2.381 |
|  |  | Property | 4446 | 4446 | 8.298 | 2.942 |
| 5 | Clinic | Object | 3780 | 3780 | 13.661 | 2.428 |
|  |  | Relationship | 22567 | 22567 | 79.020 | 2.544 |
|  |  | Property | 275 | 275 | 49.411 | 2.890 |

It should be noted that it usually takes longer for the first run of graph query because of the time needed to interpret the query command by the graph DBMS. Figure 6 shows the relationship between query time and the iterations of a query. It is obvious that the minimum time is more stable than the maximum and the mean query time. It is reasonable to claim that, for graph database, the more a query is run, the less time it requires before reaching the best query performance. The query time presented in Table 8 for Cypher4BIM is the average query time after 100 iterations.

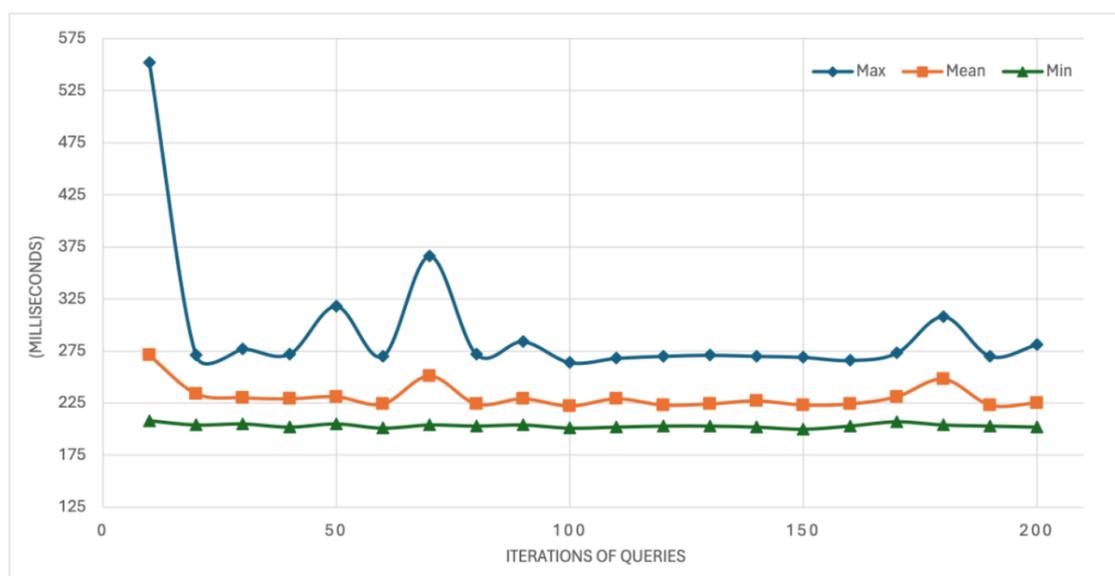

Figure 6 Relation between query time and the iterations of queries.

*4.3 Querying relations of entities*

Relations between entities are substantial part of IFC [15]. IFC4 has defined 53 relationship entities, and the common relations include spatial structure, spatial containment, space boundary, space accessibility, element connectivity, and property sets. These relations were queried using the patterns listed in Table 5, with a modification to the pattern for property set, where only the property set of the IfcProject instance was retrieved using (n: IfcProject) − −(: IfcRelDefinesByProperties) − −(: IfcPropertySet). The query results are presented in Table 9.

Table 9 Query results of relations.

|   | Model | Information | Query Results | | Query Time (seconds) | | |
|---|---|---|---|---|---|---|---|
|   |   |   | Nodes | Edges | Max | Mean | Min |
| 1 | Duplex | Spatial Structure | 33 | 64 | 0.273 | 0.230 | 0.210 |
|   |   | Spatial Containment | 154 | 300 | 0.276 | 0.150 | 0.125 |
|   |   | Space Boundary | 356 | 948 | 0.285 | 0.247 | 0.213 |
|   |   | Space Accessibility | 60 | 240 | 0.129 | 0.044 | 0.032 |
|   |   | Connectivity | 139 | 328 | 0.385 | 0.256 | 0.211 |
|   |   | Property Set | 0 | 0 | 0.223 | 0.029 | 0.005 |
| 2 | House | Spatial Structure | 16 | 30 | 0.650 | 0.113 | 0.050 |
|   |   | Spatial Containment | 86 | 168 | 0.287 | 0.097 | 0.070 |
|   |   | Space Boundary | 110 | 280 | 0.149 | 0.075 | 0.062 |
|   |   | Space Accessibility | 28 | 56 | 0.154 | 0.032 | 0.017 |
|   |   | Connectivity | 29 | 64 | 0.109 | 0.051 | 0.040 |
|   |   | Property Set | 5 | 8 | 0.018 | 0.007 | 0.005 |
| 3 | Smiley | Spatial Structure | 154 | 306 | 5.486 | 5.097 | 4.910 |
|   |   | Spatial Containment | 841 | 1672 | 1.013 | 0.862 | 0.722 |
|   |   | Space Boundary | 2273 | 6264 | 1.856 | 1.682 | 1.531 |
|   |   | Space Accessibility | 643 | 2552 | 0.447 | 0.366 | 0.292 |
|   |   | Connectivity | 639 | 1476 | 1.448 | 1.149 | 1.044 |
|   |   | Property Set | 0 | 0 | 0.218 | 0.026 | 0.004 |
| 4 | Institute | Spatial Structure | 97 | 192 | 2.813 | 1.857 | 1.601 |
|   |   | Spatial Containment | 458 | 906 | 0.613 | 0.478 | 0.381 |
|   |   | Space Boundary | 1504 | 7936 | 1.159 | 1.086 | 1.013 |
|   |   | Space Accessibility | 309 | 612 | 0.273 | 0.179 | 0.141 |
|   |   | Connectivity | 337 | 864 | 0.777 | 0.669 | 0.534 |
|   |   | Property Set | 3 | 4 | 0.063 | 0.010 | 0.003 |
| 5 | Clinic | Spatial Structure | 281 | 560 | 51.628 | 37.944 | 33.546 |
|   |   | Spatial Containment | 1682 | 3356 | 1.863 | 1.662 | 1.413 |
|   |   | Space Boundary | 4424 | 11660 | 3.707 | 3.188 | 2.936 |
|   |   | Space Accessibility | 957 | 1920 | 0.610 | 0.521 | 0.430 |
|   |   | Connectivity | 3181 | 8344 | 6.457 | 6.239 | 6.055 |
|   |   | Property Set | 0 | 0 | 0.124 | 0.019 | 0.006 |

The queried graphs from the house model and the clinic model are presented in Figure 7 and Figure 8, respectively, as examples of results. To retrieve the same amount of information using IfcOpenShell, much more work would be required, due to the manual traversal of the query path. This has been further demonstrated in the discussion section.

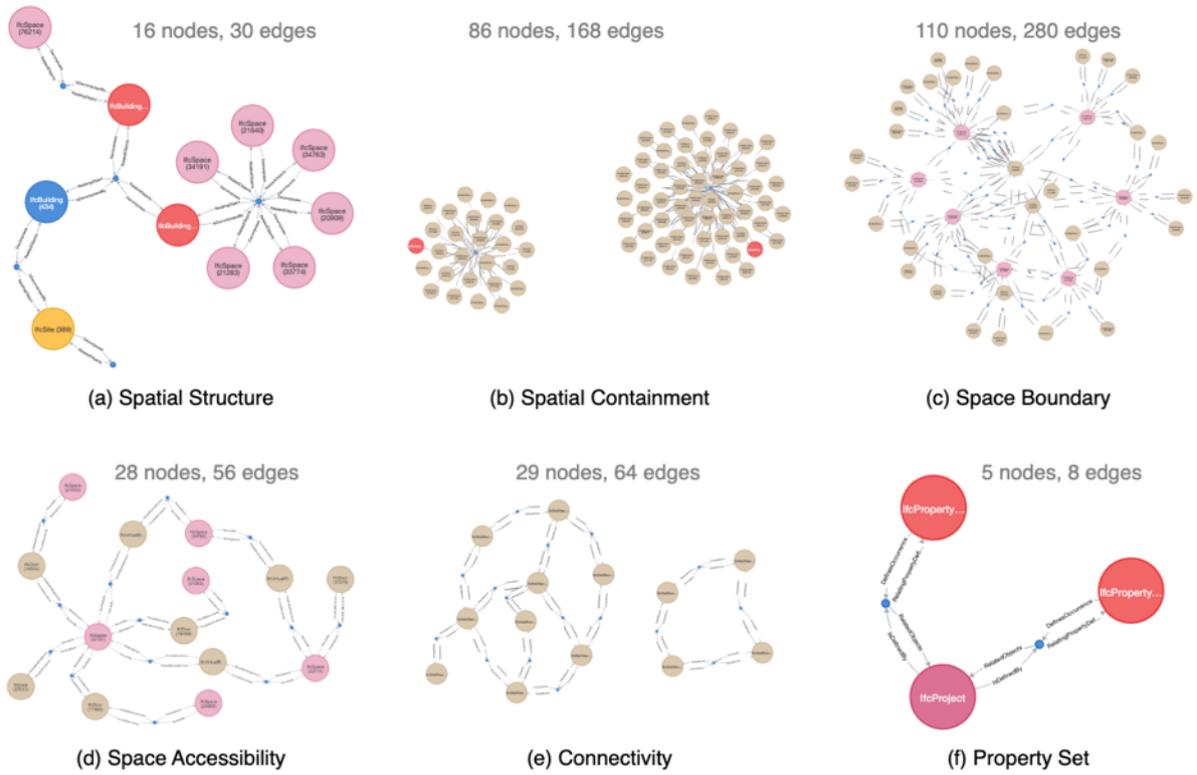

Figure 7 The queried relation graphs from the house model.

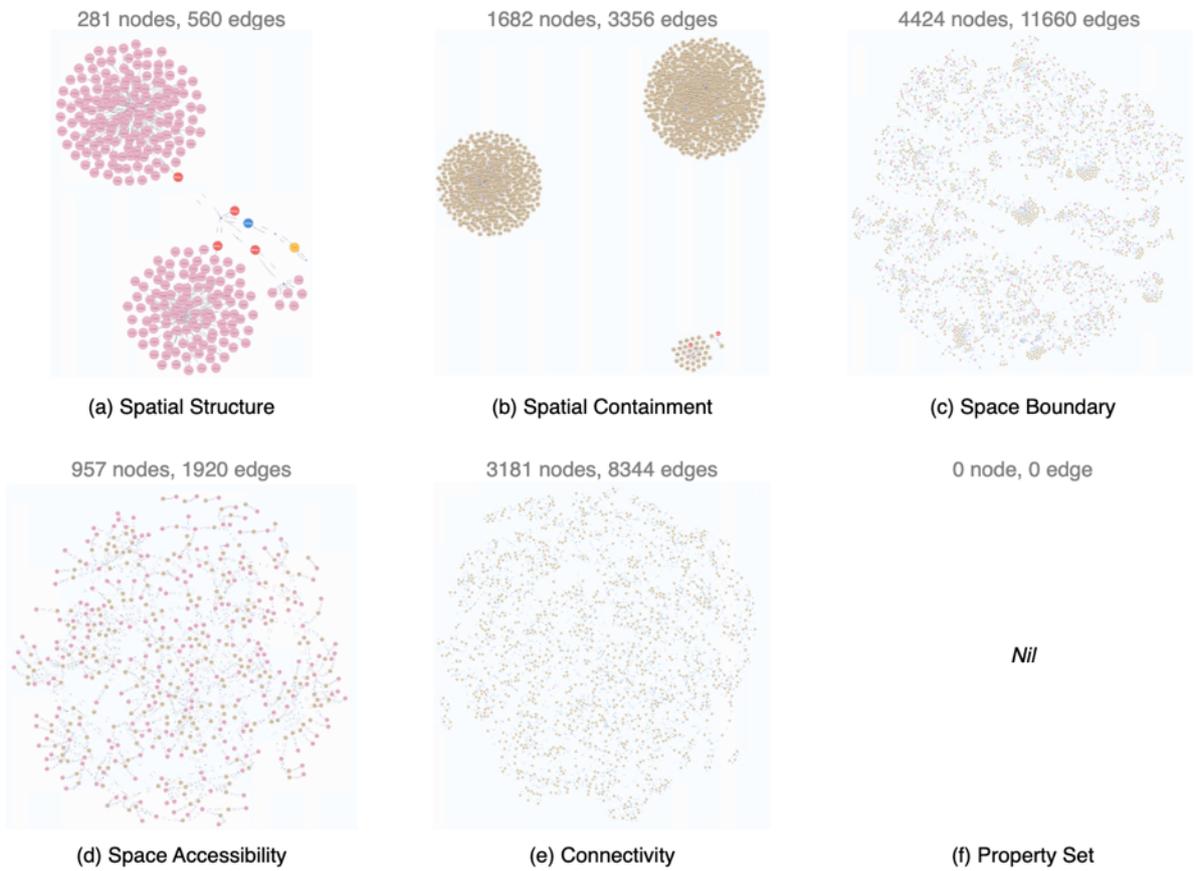

Figure 8 The queried relation graphs from the clinic model.

Most of the queries were completed in seconds, but it depends on the size of graph queried. A correlation analysis was carried out between the graph size (i.e., number of nodes and edges) and the minimum query times for queries. The result is presented in Table 10, which shows that the time needed to query these relations has a positive correlation with the number of nodes and edges, especially the number of edges.

Table 10 Correlation between graph size (nodes and edges) and query efficiency (time).

|  | Nodes | Edges |
| --- | --- | --- |
| Spatial Structure | 0.68 | 0.89 |
| Spatial Containment | 0.77 | 0.91 |
| Space Boundary | 0.82 | 0.93 |
| Space Accessibility | 0.77 | 0.86 |
| Connectivity | 0.69 | 0.90 |

## 5 Discussion

This study has revealed the power of graph and the capability of Cypher4BIM in building information query. IFC-Graph, together with Cypher4BIM, can reveal and visualise hidden information in digital building models by using graph pattern matching, which can be used for graph-based knowledge discovery [58].

### 5.1 The power of IFC-Graph/Cypher4BIM in revealing hidden information

By using pattern matching, Cypher4BIM can reveal hidden information in IFC. Such information cannot be visualised in BIM authoring tools, such Autodesk Revit. One of the examples is the boundary of space (see Figure 9). When querying the space boundary of the house model, it was noticed that a slab (59290) is connected to a space (76214) via 8 instances of IfcRelSpaceBoundary, i.e., the blue dots in Figure 9 (a).

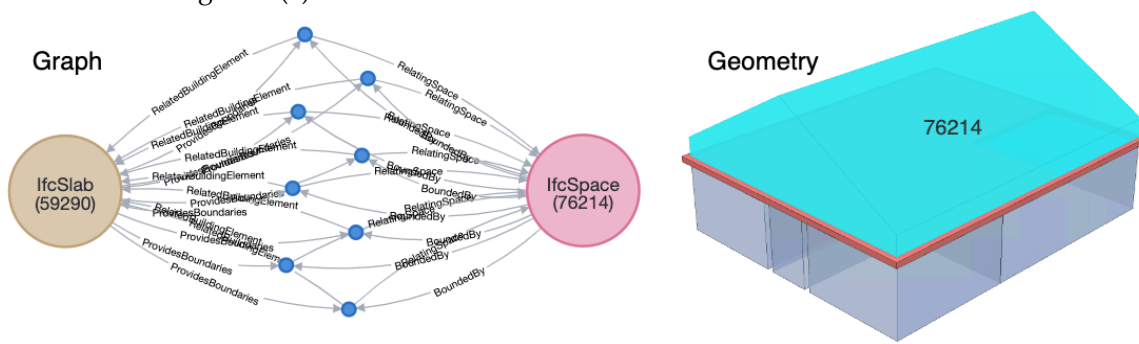

(a) Relationship between slab (59290) and space (76214)

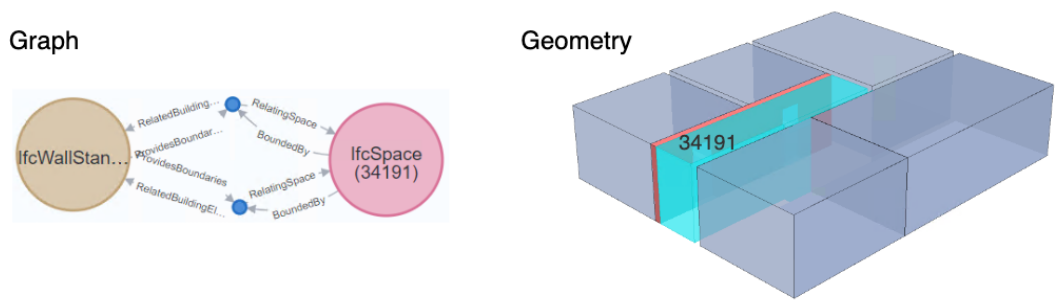

(b) Relationship between wall (18698) and space (34191)

Figure 9 Relationship between boundary elements and space.

This pattern may appear unusual in the first sight, as it does not make sense to link a single slab 8 times to a space or link a wall twice to a space (see Figure 9 (b)). It is, however, compliant with the IFC standard when space boundaries are defined at the second level. In IFC, space boundaries can be at the first level (1st level space boundary) or the second level (2nd level space boundary). The difference between them can be found in [57].

The hidden information can be vital for some applications, such as the integration of BIM and geographical information system (GIS) [2,5,59,60]. For example, the boundary information can be the key to solving a challenge in IFC-to-CityGML conversion when generating internal surfaces for walls, ceilings, floors, windows, and doors at the fourth level of detail (LoD4). LoD4 in CityGML 2.0 deals with both internal and external surfaces. Even though LoD4 has been removed from CityGML 3.0 [61], the challenge of generating internal surfaces remains.

The hidden information will be difficult to reveal without graph and pattern matching. This capability of graph in visualising hidden information makes IFC-Graph suitable for educational purpose as well where building information is involved, for example, showing the topology of building elements, as discussed by March and Steadman [62].

*5.2 The power of pattern in graph query*

The importance of pattern in graph query has been demonstrated. To further explore the power of pattern in graph query, additional work has been carried out to show how a slight change in pattern can greatly improve the effectiveness of information query (i.e., finding more information) or how the same amount of information can be retrieved using a simpler pattern.

Table 5 shows the pattern for retrieving property sets of building elements, $(n: \text{IfcBuildingElement}) - -(: \text{IfcRelDefinesByProperties}) - -(: \text{IfcPropertySet})$. With this pattern, property sets of building elements can be retrieved. However, the detailed values of properties are separately stored in individual IfcProperty instances. If IfcOpenShell is used to extract these properties, much more work would be required. However, this can be realised in a much easier way with graph and pattern. The IfcProperty instances can be retrieved by adding two empty nodes at the end of the original query pattern, making it $(n: \text{IfcBuildingElement}) - -(: \text{IfcRelDefinesByProperties}) - -(: \text{IfcPropertySet}) - -(\ ) - -(\ )$. Figure 10 shows the query results of these two patterns using the IfcProject instance as example. This slight change in pattern can be very useful when processing property sets of any IfcObject instances.

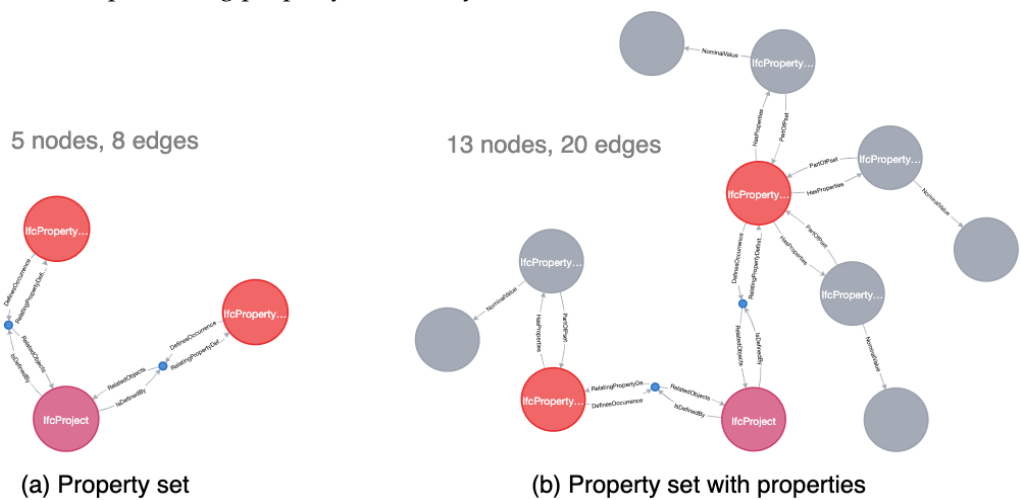

Figure 10 The effect of changing pattern in the query results.

It is possible to use a simpler pattern to retrieve the same amount of information. The spatial structure is used to illustrate this point. The spatial structure in IFC is IfcSite → IfcBuilding → IfcBuildingStore → IfcSpace, these entities are connected via IfcRelAggregates. Even though IfcProject is not part of the spatial structure, but IfcSite is linked to IfcProject via IfcRelAggregates as well. The pattern in Listing 2 (a) can be used to get the spatial structure, but it is over complex and can be simplified to be ( ) − −(: IfcRelAggregates) − −(: IfcSpatialStructure). Such a change in pattern also involves the use of the concept of abstraction, which is discussed in the next section. Therefore, the query pattern should be an important consideration when designing queries.

Listing 2 The query pattern for spatial structure.

| | |
|---|---|
| (a) | ( )-[:IsDecomposedBy]-(:IfcRelAggregates)-[:RelatedObjects]-(:IfcSite)-[:IsDecomposedBy]-(:IfcRelAggregates)-[:RelatedObjects]-(:IfcBuilding)-[:IsDecomposedBy]-(:IfcRelAggregates)-[:RelatedObjects]-(:IfcBuildingStorey)-[:IsDecomposedBy]-(:IfcRelAggregates)-[:RelatedObjects]-(:IfcSpace) |
| (b) | ( ) − −(: IfcRelAggregates) − −(: IfcSpatialStructure) |

*5.3 The power of abstraction in building information query*

Another interesting finding of this study is that the use of abstraction can greatly improve the effectiveness of information query. Abstraction is commonly used in programming and data modelling, including IFC. IFC has a hierarchical structure for its entities, which can show the parent-child relationship between entities. Child entities inherit the attributes of their parent entity and have their own attributes; therefore, child entities have more attributes and are more specific than their parent entities. Table 11 shows an example of the hierarchical structure using IfcDoor [63], where the root entity, IfcRoot, has only 4 attributes, while IfcDoor has 37 attributes in total, including 32 attributes from its parents and 5 new attributes.

Table 11 Levels of abstraction in IFC, using IfcDoor as example.

| Level | Entity | Attributes (Intrinsic & Extrinsic) | New Attributes |
|---|---|---|---|
| 1 | IfcRoot | 4 | - |
| 2 | IfcObjectDefinition | 11 | 7 |
| 3 | IfcObject | 16 | 5 |
| 4 | IfcProduct | 19 | 3 |
| 5 | IfcElement | 32 | 13 |
| 6 | IfcBuildingElement | 32 | 0 |
| 7 | IfcDoor | 37 | 5 |

This structure represents levels of abstraction of entities. A parent entity has a higher level of abstraction than a child entity. As the instances of child entities are also instances of the parent entity, parent entities can be used to retrieve more instances (or all instances of its child entities). The joint use of abstraction and pattern matching can move information query to the next level, as preliminarily demonstrated above. Another example for the joint use of abstraction and pattern matching is the retrieval of space boundary.

It is well-known that a space is surrounded by bounding building elements, including slabs, windows, doors, and walls. Therefore, to query the bounding building elements, the pattern listed in Listing 3 (a) was initially developed. Abstraction can be applied to further simplify this pattern, as listed in Listing 3 (b), to be (: IfcSpace) − −(: IfcRelSpaceBoundary) − −(: IfcBuildingElement), because slab, window, door, and wall are all child entities of building element. Both patterns can retrieve the spatial structure of building models. This case also involves the use of entity type as a constraint for query. Using constraints is a useful way for filtering information in query.

Listing 3 Patterns for querying space boundary

| | |
|---|---|
| (a) | (: IfcSpace) − −(: IfcRelSpaceBoundary) − −(: IfcSlab) OR (: IfcSpace) − −(: IfcRelSpaceBoundary) − −(: IfcWall) OR (: IfcSpace) − −(: IfcRelSpaceBoundary) − −(: IfcWindow) OR (: IfcSpace) − −(: IfcRelSpaceBoundary) − −(: IfcDoor) |
| (b) | (: IfcSpace) − −(: IfcRelSpaceBoundary) − −(: IfcBuildingElement) |

In the functional query pattern, i.e., $(s)--(r)--(o)$, constraints can be applied to any of the three nodes $(s), (r), (o)$. The more abstract the entities, the more expressive the pattern. In an extreme case, the pattern, $()--()--()$, will return the entire graph. In Table 12, the query of the spatial structure of the house model is used as an example of applying constraints to $(s)$. By removing the constraint (i.e., : IfcSpatialStructure) from $(s)$, the IfcProject instance can be additionally retrieved.

Table 12 The use of entity type as constraints to (s).

| | Pattern | Graph (for information only) |
|---|---|---|
| 1 | $(: IfcSpatialStructure)--(: IfcRelAggregates)--(: IfcSpatialStructure)$ | 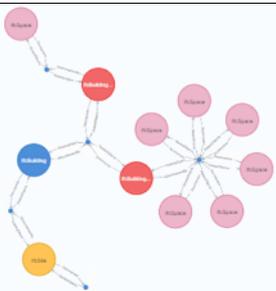 |
| 2 | $()--(: IfcRelAggregates)--(: IfcSpatialStructure)$ | 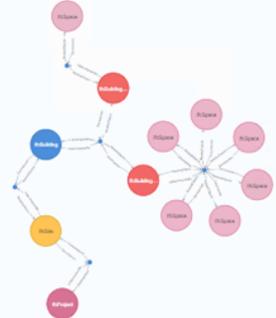 |

In another case, IfcRelSpaceBounday is used as an example of applying type constraints to $(r)$ and $(o)$. The entity hierarchy for IfcRelSpaceBounday is IfcRelationship → IfcRelConnects → IfcRelSpaceBounday. The result is presented in Table 13, including the number of nodes and edges retrieved and the time for the query. It is obvious that the more abstract the entities, the more nodes and edges retrieved by the query, but more computational resources (time) are required.

Table 13 The use of entity type as constraints to (r) and (o).

| | Pattern | Nodes | Edges | Time (s) | Graph |
|---|---|---|---|---|---|
| 1 | $()--(: IfcRelSpaceBoundary)--(: IfcWall)$ | 76 | 140 | 0.066 | 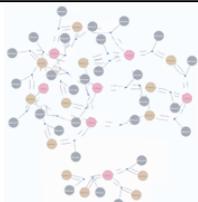 |
| 2 | $()--(: IfcRelSpaceBoundary)--()$ | 205 | 395 | 0.357 | 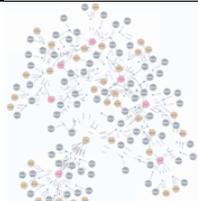 |
| 3 | $()--(: IfcRelConnects)--()$ | 320 | 719 | 4.741 | 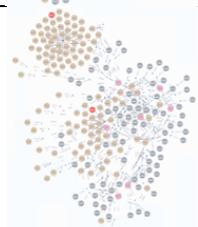 |

| | Pattern | Nodes | Edges | Time (s) | Graph |
|---|---|---|---|---|---|
| 4 | ( ) − −(: IfcRelationship) − −( ) | 1390 | 3173 | 9.608 | 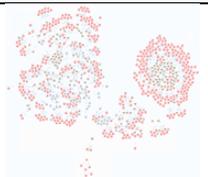 |

This study can also benefit previous studies. To use graph algorithms, previous studies extracted data from IFC-SPF and converted the extracted data into graph. A challenging step in this process is the extraction of data from IFC-SPF. In this sense, this study can benefit those studies by providing a better way for information extraction from IFC. For instance, the query in Table 13 (4) can be used to retrieve the Relation Sub Graphs required by [20] in a much quicker and easier way.

*5.4 BIM model quality check*

Cypher4BIM is built on Cypher, its patterns can be used to represent human knowledge and used in many applications, such as in rule-based compliance check and model quality check for building models [41]. The use of Cypher4BIM in model quality check is shown below using the duplex model. Modelling errors were noticed when querying the space accessibility of the duplex model, as shown in Figure 11.

From the geometric model (see Figure 11 (a)), it can be observed that door (125) is meant to connect space (152) and space (118) only. However, due to modelling error, door (125) was mistakenly placed, which resulted in two consequences. First, door (125) was thought to be connected to space (84) as well, and second, door (125) is connected to space (84), space (152), and space (116) twice, as shown in Figure 11 (b). Such a modelling error is not easy to identity. The correct graph should be the one presented in Figure 11 (c).

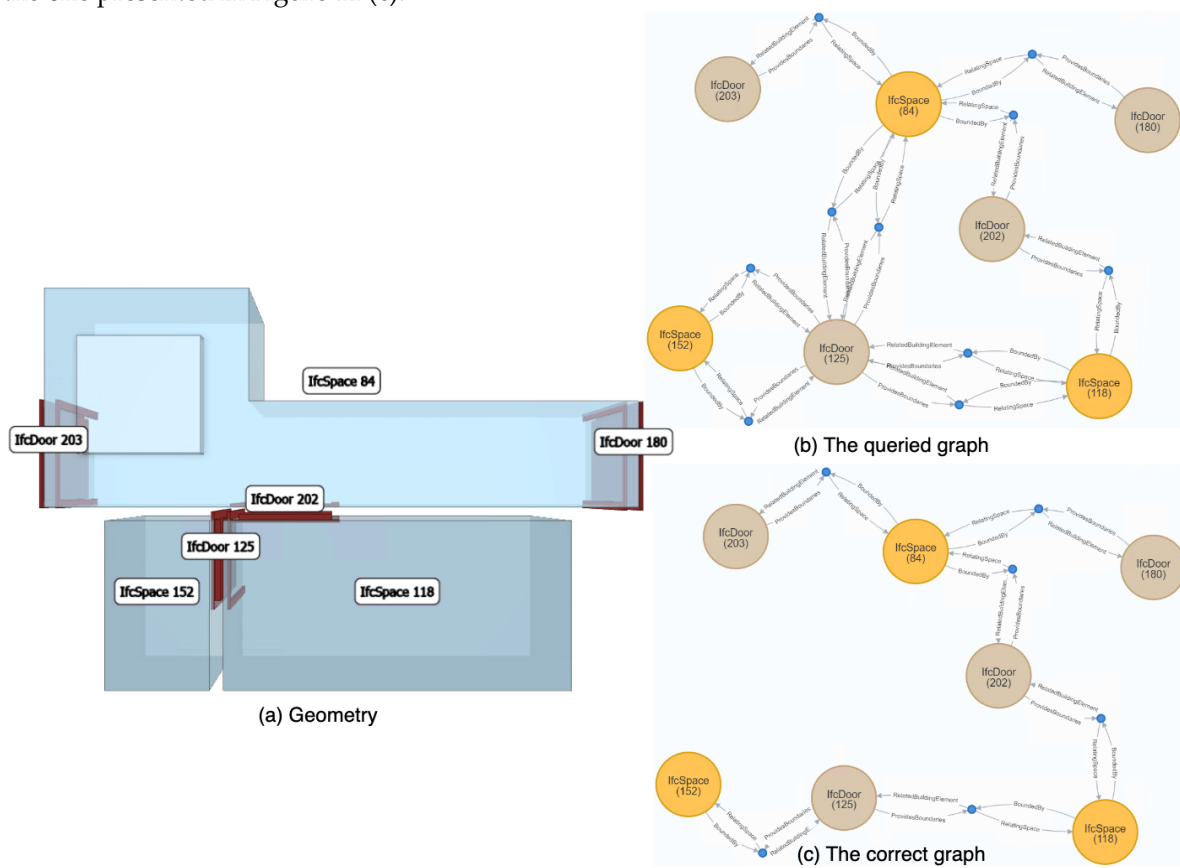

Figure 11 Modelling errors in the duplex model identified by using IFC-Graph/Cypher4BIM.

*5.5 The expressiveness of graph pattern*

To further demonstrate the expressiveness of graph pattern, a simple comparison was carried out. Table 14 shows the codes needed to retrieve the spatial structure by using Cypher4BIM, IfcOpenshell, SPARQL, and Cypher from IFC-Graph, IFC-SPF, RDF in graph form and RDF in Turtle, respectively. It is obvious that codes using graph pattern matching are more compact, thus more expressive. This is even the case for RDF. With Cypher4BIM and IFC-Graph, useful information (e.g., connectivity and accessibility) can be easily and effectively retrieved, rather than writing long queries using IFC parsers. An additional example of querying IFC-XML using XPath is also provided in Table 14 for reference.

Table 14 Codes needed to retrieve the spatial structure.

| | Format | Language | Query |
|---|---|---|---|
| 1 | IFC-Graph | Cypher4BIM | # Get all sites, buildings, storeys, spaces, and their relationship<br>match p = ( ) − −(: IfcRelAggregates) − −( ) return p |
| 2 | IFC-SPF | IfcOpenShell | # Only get the first site, the first building, the first building storey, the first space, and their relationships.<br>project = ifc_file.by_type('IfcProject')[0]<br>site_1 = project.IsDecomposedBy[0].RelatedObjects[0]<br>building_1 = site_1.IsDecomposedBy[0].RelatedObjects[0]<br>storey_1 = building_1.IsDecomposedBy[0].RelatedObjects[0]<br>space_1 = storey_1.IsDecomposedBy[0].RelatedObjects[0] |
| 3 | RDF (Graph) | Cypher | match p = () − −(r) − −()<br>where r. uri contains 'IfcRelAggregates'<br>return p |
| 4 | RDF (Turtle) | SPARQL | PREFIX ifcowl:<http://ifcowl.openbimstandards.org/IFC4#><br>Select ?Project ?Site ?Building ?BuildingStorey ?Space<br>WHERE {<br>　　?Project a ifcowl:IfcProject<br>　　?Relaggregate1 relatingObject_IfcRelAggregates ?Project<br>　　?Relaggregate1 relatedObjects_IfcRelAggregates ?Site<br>　　?Relaggregate2 relatingObject_IfcRelAggregates ?Site<br>　　?Relaggregate2 relatedObjects_IfcRelAggregates ?Building<br>　　?Relaggregate3 relatingObject_IfcRelAggregates ?Building<br>　　?Relaggregate3 relatedObjects_IfcRelAggregates ?BuildingStorey<br>　　?Relaggregate4 relatingObject_IfcRelAggregates ?BuildingStorey<br>　　?Relaggregate4 relatedObjects_IfcRelAggregates ?Space} |
| 5 | IFC-XML | XPath | //IfcRelAggregates/RelatedObjects/*[name()='IfcSite' or name()='IfcBuilding' or name()='IfcBuildingStorey' or name()='IfcSpace']/concat(../../RelatingObject/*/@ref,' , ', @ref) |

This paper does not mean to diminish IfcOpenShell. As the most used IFC-SPF parser, it has greatly contributed to research in openBIM. This paper is intended to demonstrate a potential form of building information in the future. Table 15 presents a more complicated use case, i.e., finding rooms adjacent to the external environment, which can be used in acoustic assessments.

Table 15 Using Cypher4BIM and IfcOpenShell to find rooms adjacent to external environment.

| | Language | Finding rooms adjacent to external environment for acoustic assessment |
|---|---|---|
| 1 | Cypher4BIM | match (n:IfcSpace)--(r:IfcRelSpaceBoundary)<br>where r.InternalOrExternalBoundary = 'EXTERNAL'<br>return n |
| | Result | IfcSpace (21640), IfcSpace (33774), IfcSpace (34191), IfcSpace (21283), IfcSpace (34763), IfcSpace (20909), IfcSpace (76214) |
| 2 | IfcOpenShell | def find_rooms_adjacent_to_external(ifc_file):<br>    def Is_external_boundary(boundaries):<br>        flag = False<br>        for boundary in boundaries:<br>            if boundary.InternalOrExternalBoundary == 'EXTERNAL':<br>                flag = True<br>        return flag<br>    rooms = ifc_file.by_type('IfcSpace')<br>    rooms_external = []<br>    for room in rooms:<br>        boundaries = room.BoundedBy<br>        if Is_external_boundary(boundaries):<br>            rooms_external.append(room)<br>    return rooms_external |
| | Result | #20909=IfcSpace('347jFE2yX7IhCEIALmupEH',#12,'4',$,$,#20819,#20904,'Schlafzimmer',.ELEMENT.,$,$)<br>#21283=IfcSpace('0e_hbkIQ5DMQlIJ$2V3j_m',#12,'3',$,$,#21203,#21278,'Bad',.ELEMENT.,$,$)<br>#21640=IfcSpace('2RSCzLOBz4FAK$_wE8VckM',#12,'2',$,$,#21560,#21635,'Buero',.ELEMENT.,$,$)<br>#33774=IfcSpace('0Lt8gR_E9ESeGH5uY_g9e9',#12,'5',$,$,#33683,#33769,'Wohnen',.ELEMENT.,$,$)<br>#34191=IfcSpace('3$f2p7VyLB7eox67SA_zKE',#12,'1',$,$,#34051,#34186,'Flur',.ELEMENT.,$,$),<br>#34763=IfcSpace('17JZcMFrf5tOftUTidA0d3',#12,'6',$,$,#34672,#34758,'Küche',.ELEMENT.,$,$),<br>#76214=IfcSpace('2dQFggKBb1fOc1CqZDIDlx',#12,'7',$,$,#76123,#76209,'Galerie',.ELEMENT.,$,$) |

*5.6 Limitation and furture work*

    Even though this study has demonstrated many features of IFC-Graph and Cypher4BIM, some limitations were noticed as well regarding the usability of the queried information and the simplification of graph.

    1) The usability of the queried information. Cypher4BIM is mainly for querying the raw information in IFC-Graph, including the semantic information and the raw geometric information that is for computing explicit geometry. To make such raw information more practical and usable,

further processing is required. For example, using the pattern for space accessibility query, a graph showing space connectivity can be obtained. However, such a graph cannot be directly used in applications, such as route planning for emergency response [64], because the queried space connectivity is for individual floors only (see Figure 7 (e)). Additional processing is required to establish connections between floors to eventually produce a whole-building connectivity graph. At this moment, this probably requires the use of geometric information from stairs and their relationship with storeys. But in the future, if an objectified relationship can be created by buildingSMART to show the connection between stairs and spaces, it would be much easier to get the whole-building connectivity graph, without the need to compute explicit geometry.

2) Modification/Simplification of graph. To make graphs even more practical, additional work is required to simplify the graph. This can be carried out in two ways. First, eliminating unnecessary attributes. This is because extrinsic attributes of entities, such as 'RelatedBuildingElement' and 'RelatingSpace', are barely used in query. Second, replacing some relationship entities with attributes. According to van Berlo [15], where possible, objectified relationships should be replaced with direct attributes, as presented in Figure 12. This is possible when entities are connected by 1:1 relationship. The simplification of IFC-Graph can be thought of as developing a new data model. Simplification of graph is a good way to reduce graph size, improve efficiency of query, and further improve the usability of graphs in practice. It should be noted that simplification of graph is different from simplification of the IFC data model. The simplified graphs actually represent new data models tailored for certain use cases.

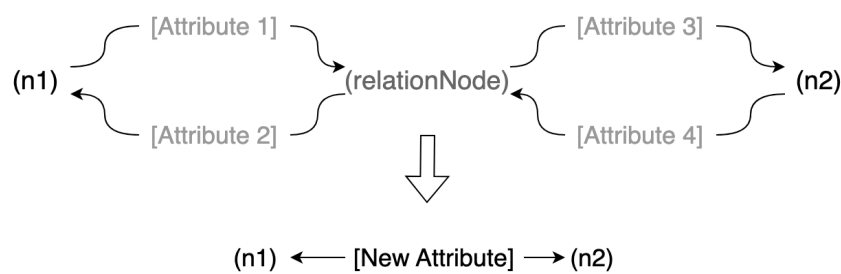

Figure 12 Modifying IFC-Graph for better usability.

## 6 Conclusion

Graph has been considered a highly potential and new form for building information storage and query. This study tries to understand how to effectively query IFC information from IFC-Graph, especially the hidden relation information and develop a graph-based language for building information query. The developed query language is based on Cypher and referred to as Cyphter4BIM. This query language is mainly for application programmers and sophisticated users who have the strong need to retrieve and use building information from IFC in a more effective and efficient manner. The main findings and outcomes of this study are as follows.

(1) The graphic representation of building information (IFC-Graph) can make full use of graph algorithms to retrieve building information from interconnected datasets, and graph database has better performance for information query than file-processing parsers, especially for retrieving relations. (2) A set of advanced functional query patterns, which are based on but different from the basic query patterns, have been developed for querying individual instances and relations from IFC-Graph, such as spatial structure, space boundary, space accessibility, element connectivity, property sets, and quantity sets. (3) It is noticed that the joint use of pattern and the concept of abstraction can be very useful in improving the efficiency of information query, and constraints, such as entity type, can be applied to any nodes in a path to limit the query results.

Graph database management systems provide high interoperability between digital twin systems through the provision of API and have an excellent support for Artificial Intelligence (AI), therefore, this study contributes to the development of digital twin and graph-based (AI) as well in a general sense. Further work is required to make the queried graph more usable in practice, for example, by simplifying graph through eliminating attributes and some objectified relationships.

**Acknowledgments:** This project has received funding from the European Union's Horizon 2020 research and innovation programme under the Marie Skłodowska-Curie grant agreement No 101034337. We would like to thank Dr. Lavindra De Silva, Dr. Pieter Pauwels, Dr. Stefano Cavazzi, Yogesh Patel, Matt Peck, Katrin Johannesdottir, George Economides, Ajay Gupta, Peter Lindgren, and Alex Gillies for their contributions to this paper. The authors would like to thank the anonymous reviewers for their comments and suggestions that helped improve the comprehensiveness and clarity of our paper.

**Author Contributions:** Junxiang Zhu, Nicholas Nisbet, Ran Wei, and Ioannis Brilakis conceived of and designed the experiments; Junxiang Zhu and Mengtian Yin performed the experiments and analysed the data; Junxiang Zhu and Ioannis Brilakis wrote the paper. Junxiang Zhu revised the paper.

**Conflicts of Interest:** The authors declare no conflicts of interest.

**Appendix**

Table 16 Python codes for testing query time of IfcOpenShell for individual instances.

```
models = ['AC20_FZK_Haus.ifc','AC20-Institute-Var-2.ifc',\
    'AC-20-Smiley-West-10-Bldg-modified-for-neo.ifc',\
    'Duplex_A_20110907_optimized.ifc','Clinic_A_20110906_optimized.ifc']
for model in models:
   print(model)
   f = ifcopenshell.open('./data/'+model)
   query_time=[]
   for i in range(x):
      tic()
      f.by_type('IfcRelationship')
      t = toc()
      query_time.append(t)
   df = pd.DataFrame(query_time)
   print([1000*df.max(),1000*df.mean(),1000*df.min()])
   print('------')
```

Table 17 Python codes for testing query time of Cypher4BIM for individual instances and relation.

```
def test_object (graph):
    cypher = 'MATCH (n:IfcObjectDefinition) return count(n)'
    result = graph.run(cypher)
    return result

def get_spatial_structure(graph):
    cypher = 'MATCH p=(n1:IfcSpatialStructureElement)--(r:IfcRelAggregates)--(n2:IfcSpatialStructureElement) RETURN p'
    result = graph.run(cypher)
    return result

query_time=[]
for i in range(x):
    tic()
    test_object(graph_1)
    t = toc()
    query_time.append(t)
df = pd.DataFrame(query_time)
[1000*df.max(),1000*df.mean(),1000*df.min()]
```